\documentclass[%
 reprint,
 amsmath,amssymb,
 aps, pre
]{revtex4-1}
\usepackage{multirow}
\usepackage{tikz}
\usepackage{array}
\usepackage{subcaption}
\usepackage{caption}
\usepackage{graphicx}
\usepackage{dcolumn}
\usepackage{bm}
\usepackage[colorlinks=true, allcolors=blue]{hyperref}

\begin{document}

\preprint{APS/123-QED}

\title{Causality Testing: A Data Compression Framework}

\author{Aditi Kathpalia}
 \email{kathpaliaaditi@gmail.com} 
\author{Nithin Nagaraj}%
\email{nithin@nias.res.in}
\affiliation{%
National Institute of Advanced Studies, \\Indian Institute of Science Campus, Bengaluru 560012, India\\
}%




\date{\today}

\begin{abstract}

Causality testing, the act of determining cause and effect from measurements, is widely used in physics, climatology, neuroscience, econometrics and other disciplines. As a result, a large number of causality testing methods based on various principles have been developed. Causal relationships in complex systems are typically accompanied by entropic exchanges which are encoded in patterns of dynamical measurements. A data compression algorithm which can extract these encoded patterns could be used for inferring these relations. This motivates us to propose, for the first time, a generic causality testing framework based on data compression. The framework unifies existing causality testing methods and enables us to innovate a novel Compression-Complexity Causality measure. This measure is rigorously tested on simulated and real-world time series and is found to overcome the limitations of Granger Causality and Transfer Entropy, especially for noisy and non-synchronous measurements. Additionally, it gives insight on the `kind' of causal influence between input time series by the notions of positive and negative causality.

\begin{description}
\item[Keywords]
Causality testing, data compression, compression-complexity, negative causality.
\item[PACS numbers]89.70.Cf, 87.19.lo, 05.45.Tp, 89.75.-k
\end{description}
\end{abstract}

\pacs{Valid PACS appear here}
\maketitle


\section{\label{introduction}Introduction and Framework}

Causality testing is used widely in various disciplines including neuroscience \cite{seth}, physics~\cite{crutchfield}, climatology~\cite{stips}, econometrics~\cite{chen} and epidemiology~\cite{sachs}. A number of causality testing methods exist which can be applied to estimate the magnitude and direction of causality between two time series, some of which are listed in Fig.~\ref{block_diagram}. These methods make very different model assumptions about input time series, making an appropriate choice often difficult for a given context. Thus, there is a need for a unifying framework to explain the working of these measures and suitably guide their application. 

According to Wiener~\cite{wiener}, if incorporating the past of a time series $Y$ helps to improve the prediction of a time series $X$, then $Y$ {\textit{causes}} $X$. Existing methods are based on notions of improved predictability (Granger Causality and its variations), reduction of uncertainty (Transfer Entropy, Information Flow), dynamical modeling (Dynamic Causal Modeling) and estimation based on proximity in attractor manifold (Convergent Cross Mapping) (refer Fig.~\ref{block_diagram}). All these notions are closely related to information transfer of one form or the other. By establishing the relation of information to the reality of our world, the basis of causality on information transfer is increasingly being shown to have a rigorous physical foundation \cite{ebeling,liang2}. In \cite{ebeling}, information transfer in complex systems (based on Shannon Entropy) is fundamentally linked to energy and entropy (Boltzmann) flows. In case of living and other complex systems, information is created by self-organization, resulting in a re-ordering of entropy with a reduction in their interior and a corresponding increase in the external environment. Causal relationships based on these entropic exchanges are encoded in patterns of dynamical measurements. A \lq{}data compression\rq{} framework that captures these patterns seems to be the most natural approach to extract these causal relationships. 

Inspired by decades of work in data compression~\cite{salomon}, we propose for the first time, a generic framework for causality testing. Data compression is concerned with encoding information either by way of modeling statistical redundancy (eg., Huffman coding~\cite{huffman}) or learning patterns algorithmically (eg., Lempel-Ziv coding~\cite{ziv})) with the aim of reducing resources to store or transmit data. This framework is well founded mathematically owing to a close link between data compression, information entropy (Shannon's source coding theorem~\cite{shannon}) and algorithmic complexity (Kolmogorov complexity~\cite{cover}). 

\begin{table*}
\usetikzlibrary{positioning,fit,calc,arrows,backgrounds,matrix}
\tikzset{block/.style={draw,thick,text width=2.15cm,minimum height=1cm,align=center},
        line/.style={-latex}
        comment/.style={rectangle, inner sep= 5pt, text width=2 cm, node distance=0.25cm}, 
}
\tikzstyle{input} = [coordinate]
\tikzstyle{output} = [coordinate]

\begin{tikzpicture}
  \node[block] (a) {Pre-processing/ Transform};
  \node[input,left=0.7cm of a]   (inputx) {};
  \node[block,right=0.6 cm of a] (b) {Model};
  \node[block,right=0.6 cm of b] (c) {Quantizer};
  \node[block,right=0.6 cm of c] (d) {Entropy Coding};
  \node[block,right=0.6 cm of d] (e) {Testing Criteria};
  \node[output,right=0.7cm of e]   (outputf) {};
 \draw[->]([yshift=2mm]inputx.east)-- node[left=3mm]{$X$}([yshift=2mm]a.west);
 \draw[->]([yshift=-2mm]inputx.east)-- node[left=3mm]{$Y$} node[above=8mm, text width=1.7 cm,align=left] {Input Data}([yshift=-2mm]a.west);
 \draw[->] ([yshift=2mm]a.east)-- ([yshift=2mm]b.west);
  \draw[->] ([yshift=-2mm]a.east)-- ([yshift=-2mm]b.west);
   \draw[->] ([yshift=2mm]b.east)-- ([yshift=2mm]c.west);
  \draw[->] ([yshift=-2mm]b.east)-- ([yshift=-2mm]c.west);
   \draw[->] ([yshift=2mm]c.east)-- ([yshift=2mm]d.west);
  \draw[->] ([yshift=-2mm]c.east)-- ([yshift=-2mm]d.west);
  \draw[->] ([yshift=2mm]d.east)-- ([yshift=2mm]e.west);
  \draw[->] ([yshift=-2mm]d.east)-- ([yshift=-2mm]e.west);
   \draw[->](e.east)-- node[above=0.8mm, right, text width=2.0cm, align=center] {Estimated Causality}(outputf.west);
\node[below=2.5mm of a,text width=2.5cm,align=center,font=\scriptsize\sffamily] (commenta) {Filtering, De-trending, Binning, Fourier, Wavelet, Kernel feature space, Linear/Affine, Non-Linear Transform};
\node[below=2.5mm of b,text width=2.5cm,align=center,font=\scriptsize\sffamily] (commentb) {Autoregressive,  Info-theoretic, Markov, Dictionary-based, Pattern Substitution, Dynamical model, Time delay embedding};
\node[below=2.5mm of c,text width=2.5cm,align=center,font=\scriptsize\sffamily] (commentc) {Lossy to Lossless};
\node[below=2.5mm of d,text width=2.5cm,align=center,font=\scriptsize\sffamily] (commentd) {Huffman Coding, Run Length Encoding, Arithmetic Coding, Lempel-Ziv Coding};
\node[below=2.5mm of e,text width=2.5cm,align=center,font=\scriptsize\sffamily] (commente) {F-statistic, Hypothesis testing, Entropy Rate, Cross-mapping, Bayesian criteria, Compression-Complexity Rate};

\end{tikzpicture}

\renewcommand{\arraystretch}{1.5}
\centering



\newcolumntype{C}[1]{>{\centering\let\newline\\\arraybackslash\hspace{0pt}}m{#1}}

\begin{tabular}{|C{0.4\textwidth}|C{0.6\textwidth}|} \hline
\textbf{Causality Methods} &\textbf{Brief Description}\\ 
\hline
Granger Causality (GC) \cite{Granger} & Autoregressive (AR) $\rightarrow$ F-statistic, Hypothesis testing \\ \hline
GC using Fourier and Wavelet Transform \cite{dhamala} & Fourier/Wavelet Transform $\rightarrow$  AR $\rightarrow$ Natural logarithm of spectral power ratios\\ \hline
Kernel (Non-linear) GC \cite{marinazzo} & Kernel (inhomogeneous polynomial or gaussian kernel) $\rightarrow$ AR $\rightarrow$ Filtered linear Granger causality index \\ \hline
Transfer Entropy (TE) \cite{schreiber} & Binning/Estimating continuous PDFs $\rightarrow$ Markov $\rightarrow$ Entropy rate \\ \hline
Dynamical Causal Modelling (DCM) \cite{friston} & Non-linear state-space model $\rightarrow$ Bayesian criteria \\ \hline
Convergent Cross Mapping (CCM) \cite{sugihara} & Time-delay embedding $\rightarrow$ Cross Mapping \\ \hline
Information Flow \cite{liang} & Differentiable vector fields $\rightarrow$ Rate of information Flow \\ \hline
{\bf Compression-Complexity Causality (CCC)} &  Binning $\rightarrow$ Pair substitution $\rightarrow$ Compression-complexity rate \\ \hline
\end{tabular}


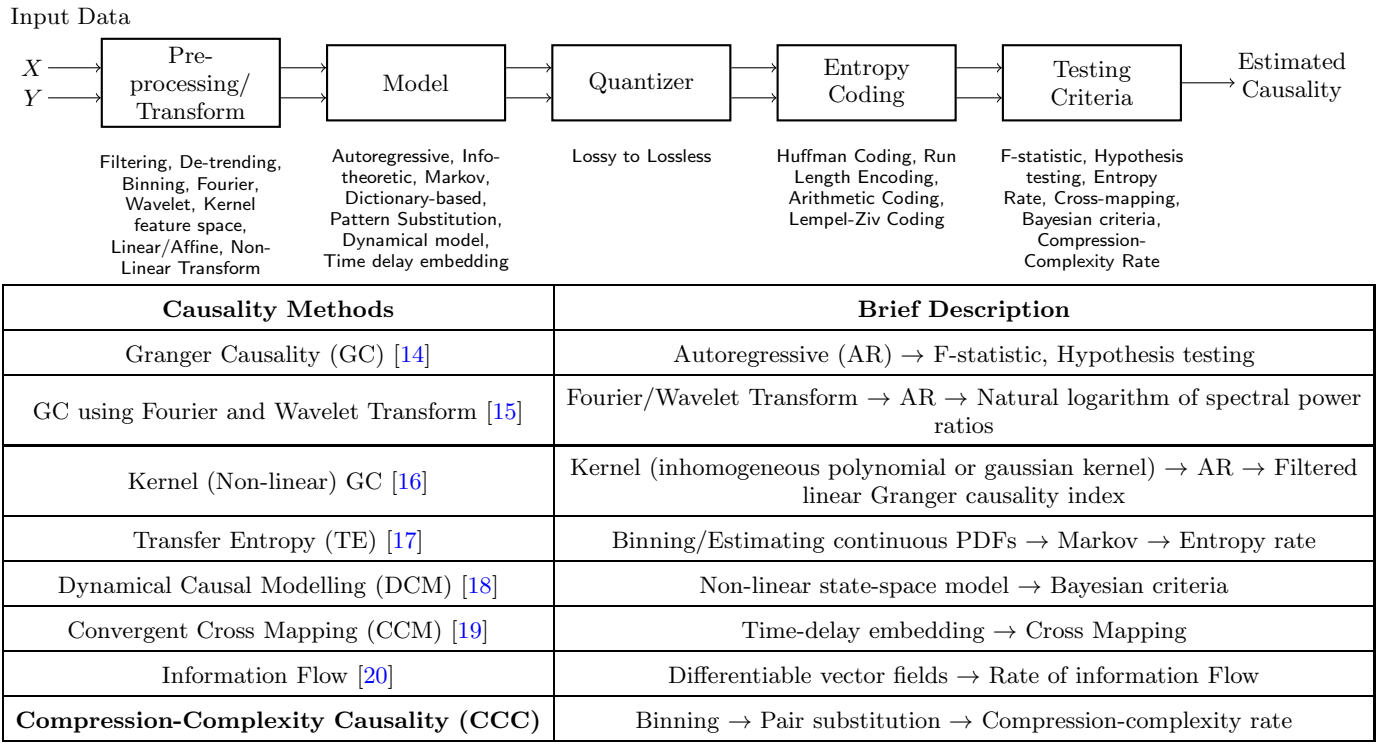
\captionof{figure}{A unifying data compression framework for causality testing. Top: Block diagram of the proposed framework indicating various stages of causality testing. Bottom: Table depicting the flow for various existing and proposed (in bold) causality testing methods.}

\label{block_diagram}

\end{table*}

Fig.1 gives the block diagram for the framework with a list of possible choices (not exhaustive) for each block below. The table describes the work-flow of various causality testing methods indicating their choice for each block. We describe three methods below as examples. Granger Causality uses only two blocks --- `Model' (autoregressive processes) and `Testing Criteria' (F-statistic). Transfer Entropy uses three blocks --- `Pre-processing' (estimating probability density functions using binning or coarse-graining, adaptive histogram, $k$-th nearest neighbor techniques~\cite{vicente2011transfer} etc.), `Model' (markov) and `Testing Criteria' (entropy rate). Convergent Cross Mapping uses two blocks --- `Model' (time delay embedding) and `Testing Criteria' (ability of the \textit{cause} variable to estimate the \textit{effect} variable using nearest neighborhood forecasting method). All these methods clearly fit within the proposed framework.

The framework makes no assumptions about the nature of data, is flexible and easily configurable for a given application by an appropriate choice for each block. Furthermore, the framework lends itself to the invention of novel methods of causality testing. In this work, we propose a  new measure --– Compression-Complexity Causality (CCC) which is described below.

\section{\label{CCC}Compression-Complexity Causality}
The minimum description length principle~\cite{rissanen} formalizes the Occam's razor and states that the best hypothesis (model and its parameters) for a given set of data is the one that leads to its best compression. Extending this principle for causality estimation, if the \textit{compressibility} of time series $X$ remains unchanged even upon incorporating information from time series $Y$, then we conclude that there is no causal influence from $Y$ to $X$ (implying that time series $Y$ has no role in modeling $X$). However, if there is a \textit{change in compressibility} of $X$ when $Y$ is included in its model, then we infer a causality from $Y$ to $X$.   

CCC uses lossless data compression algorithms, e.g. Lempel-Ziv (LZ)~\cite{ziv, lempel} and Effort-to-Compress (ETC)~\cite{nagaraj1}, to estimate compressibility using the notion of \textit{compression-complexity}~\cite{nagaraj3} (see Section 1 of Supplemental Material). While either LZ/ETC or any other compression-complexity measure could be used to compute CCC, in this work we use ETC as it has been found to perform better than LZ for short and noisy time series~\cite{nagaraj2}. The given series are first binned --- converted to a sequence of symbols using `$B$' uniformly sized bins for the application of these complexity measures. For binned time series $X$ and $Y$ of length $N$, to determine whether $Y$ causes $X$ or not, we consider a moving window $\Delta X$ of length $w$ and define compression-complexity rates as follows: 
\begin{equation}
\label{eq_complexity_1D}
CC(\Delta X \vert X_{past} )=ETC(X_{past}+\Delta X)-ETC(X_{past} ),
\end{equation}
\begin{multline}
\label{eq_complexity_2D}
CC(\Delta X \vert X_{past}, Y_{past} ) = ETC(X_{past}+\Delta X, Y_{past}+\Delta X)\\ - ETC(X_{past},Y_{past} ),
\end{multline}
where the compressibility of $\Delta X$ is estimated based on windows of immediate past $L$ values, $X_{past}$ and $Y_{past}$, taken from $X$ and $Y$ respectively. `$+$' refers to appending, for e.g., for time series $A=[1,2,3]$ and $B=[p,q]$, then $A+B=[1,2,3,p,q]$. Eq.~\ref{eq_complexity_1D} gives the compression-complexity rate defined as the effort-to-compress $\Delta X$, knowing the recent past of $X$ alone. Eq.~\ref{eq_complexity_2D} is the compression-complexity rate for $\Delta X$ knowing the recent pasts of both $X$ and $Y$. $ETC(\cdot)$ and $ETC(\cdot, \cdot)$ refer to individual and joint \textit{effort-to-compress} complexities (see Section 2 of Supplemental Material for details).

We now define Compression-Complexity Causality $CCC_{Y \rightarrow X}$ as 
\begin{equation}
\label{eq_CCC}
CCC_{Y \rightarrow X}= \overline{CC}(\Delta X \vert X_{past} ) - \overline{CC}(\Delta X \vert X_{past}, Y_{past} ),
\end{equation}
which is the difference between the two time averaged compression-complexity rates over the entire length of the time series with the window $\Delta X$ being slided by a step-size of $\delta$. If $\overline{CC}(\Delta X \vert X_{past}, Y_{past}) \approx \overline{CC}(\Delta X \vert X_{past})$, then $CCC_{Y \rightarrow X}$ is statistically zero, implying no causal influence from $Y$ to $X$. If $CCC_{Y \rightarrow X}$ is statistically significant different from zero, then we infer that $Y$ causes $X$. Higher the magnitude of $CCC_{Y \rightarrow X}$, implies higher the degree of causation. 

As shown in the table of Fig.~\ref{block_diagram}, CCC uses the following blocks --- `Pre-processing' (binning), `Model' (pair substitution) and `Testing Criteria' (compression-complexity rate).

Our formulation has a striking resemblance to Transfer Entropy~\cite{schreiber}. In fact, the terms $\overline{CC}(\Delta X \vert X_{past}, Y_{past})$ and $\overline{CC}(\Delta X \vert X_{past})$ asymptotically ($N \rightarrow \infty$) approach entropy rates used in TE formulation (see Eq.~(3) and (4) of \cite{schreiber}) for stationary ergodic processes when CC is computed using an optimal lossless data compression algorithm~\cite{shannon,cover}.

  However, there are important differences between TE and CCC. In TE, $Y$ is said to cause $X$ if there is a reduction in entropy rate of $X$ when $Y$ is included in the model of $X$. However, TE is limited by the fact that (i) the model strictly assumes Markovian property which may not be valid for the given data, (ii) entropy rate of $X$ when $Y$ is included can never increase (conditioning always reduces entropy). In contrast, for CCC, (i) the model is non-linear and generic since it is based on optimal lossless data compression algorithms, (ii) compression-complexity rate of $X$ when $Y$ is included can either decrease, indicating \textit{positive causality}, or increase, indicating \textit{negative causality} from $Y$ to $X$.
\section{\label{neg_causality}Concept of Negative Causality}
The notion of positive and negative causality, which we propose, is analogous to the concept of positive and negative correlation. It is a richer characterization of causality than discussed before in literature. Consider the following two cases. Case~(A): Minimally coupled autoregressive (AR) processes: $X(t) = aX(t-1) + cY(t-1) +\epsilon_{X,t}$, $Y(t) = bY(t-1) +\epsilon_{Y,t}$, where $a = 0.9$, $b = 0.8$, $c = 0.8$, $t=1$ to $1000$s, sampling period = 1s. Noise terms, $\epsilon_Y,\epsilon_X = \nu\eta$, where $\nu$ = noise intensity = 0.03 and $\eta$ follows standard normal distribution. Here, $CCC_{Y \rightarrow X}=0.0869$, $CCC_{X \rightarrow Y}=0.0012$ (settings: $L=150$, $w=15$, $\delta=80$, $B=2$). Case~(B): Non-linearly coupled deterministic processes: $X(t)=X(t-1)+(Y(t-2) \mod 5)$, $Y(t)=Y(t-1)+1$, $t=1$ to $1000$s, sampling period = 1s. Here, $CCC_{Y \rightarrow X}=-0.0962$, $CCC_{X \rightarrow Y}=-0.0324$ (settings: $L=80$, $w=15$, $\delta=40$, $B=8$). In both cases, $X$ and $Y$ depend linearly on their past. In (A), $X$ depends linearly on the past of $Y$ while in (B), this dependence is non-linear. For (A), the kind of information transferred from $Y$ to $X$ is such that $CC(\Delta X \vert X_{past}, Y_{past} ) < CC(\Delta X \vert X_{past})$ whereas in (B), $CC(\Delta X \vert X_{past}, Y_{past} ) > CC(\Delta X \vert X_{past})$. From this, we infer that compressibility of $\Delta X$ is reduced (as compression-complexity is increased), due to non-linear influence of $Y$ in case~(B). This happens because the kind of information brought by the past of $Y$ to compress $X$ is different from the kind of information brought by the past of $X$ itself. Here, $CCC_{Y \rightarrow X}<0$, and we say that $Y$ \textit{negatively causes} $X$. On the other hand, for case~(A), $CCC_{Y \rightarrow X}>0$, and we say that $Y$ \textit{positively causes} $X$. Now, we can not only speak of $Y$ causing $X$, but also infer the kind of information that is transferred from $Y$ to $X$ based on the sign of $CCC_{Y \rightarrow X}$. As in case (B), certain kinds of non-linear influence between two time series can lead to negative causality, but there could be other mechanisms as well.

%

\begin{figure}[h!]
\includegraphics[width=\columnwidth,trim={0 0 0 0},clip]{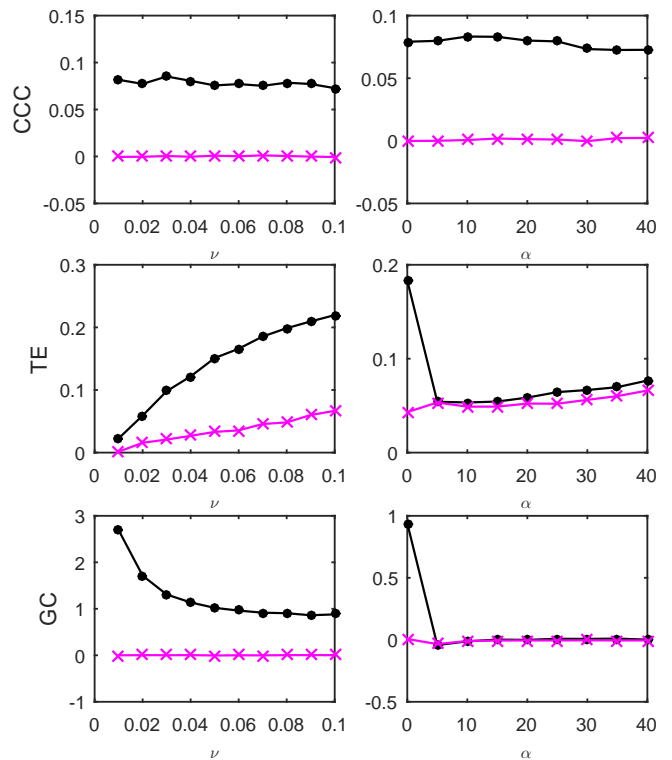}
\caption{(color online). Causality estimated using CCC, TE and GC for coupled AR processes, from Y to X (solid line-circles, black) and X to Y (dashed line-crosses, magenta) as the intensity of noise, $\nu$ (left column), and percentage of non-uniform sampling, $\alpha$ (right column), are varied.}
\label{AR_fig}
\end{figure}
\begin{figure}[t]
\includegraphics[width=\columnwidth]{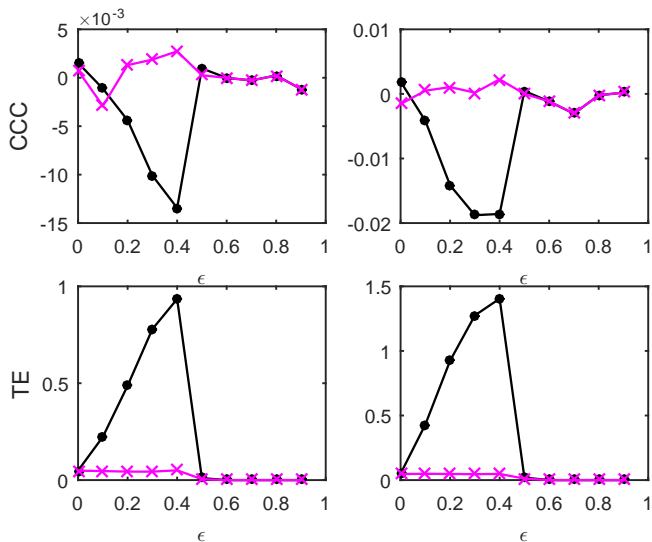}
\caption{(color online). Causality estimated using CCC and TE for coupled tent maps, from Y to X (solid line-circles, black) and X to Y (dashed line-crosses, magenta) as the degree of coupling is increased for linear coupling (left column) and non-linear coupling (right column).}
\label{tent_fig}
    \end{figure}
%
%
%

%
\section{\label{results}Results and Discussion}
\textit{Testing on simulations:} The performance of CCC was compared with that of TE and GC for the case of minimally coupled AR processes simulated as defined before. Spurious causalities using GC and TE in case of noise and low temporal resolution have been discussed in literature \cite{nalatore,hahs,smirnov}. Here, we move a step ahead and present results for non-uniformly sampled/non-synchronous measurements common in real-world physiological data acquisition due to jitters/motion-artifacts as well as in economics~\cite{baumohl}. To realistically simulate such a scenario, non-uniform sampling was introduced by eliminating data from random locations of the dependent time series and then presenting the data as a set with no knowledge of the time-stamps of the missing data. The percentage of non-uniform sampling/non-synchronous measurements ($\alpha$) is the percentage of these missing data points. Mean causality estimated for 50 trials using the three measures with increasing noise intensity ($\nu$), are shown in Fig.~\ref{AR_fig} (left column), and with increasing $\alpha$, while $\nu=0.07$, are shown in Fig.~\ref{AR_fig} (right column). Length of time series, $N=1000$, CCC settings used: $L=150$, $w=15$, $\delta=80$, $B=2$. For TE estimation, Markovian models of order 1 and $B=8$ are assumed throughout this paper. CCC estimates positive causality from $Y$ to $X$ and is statistically zero in the opposite direction. The values are stable for all cases but show a mildly increasing trend for $Y$ to $X$ causation when $\alpha$ is increased. In contrast, both TE and GC show confounding values of estimated causality in the two directions for increasing $\alpha$.

The second test involved simulation of linearly and non-linearly coupled chaotic tent maps where the independent process, $Y(t)=2Y(t-1)$ if $0 \leq Y(t-1) < 1/2$ and $Y(t)=2-2Y(t-1)$ if $1/2 \leq Y(t-1) \leq 1$. For linear coupling, the dependent process, $X(t)=\epsilon Y(t-1)+(1-\epsilon)h(t)$ where $h(t)=2X(t-1)$ if $0 \leq X(t-1) < 1/2$ and $h(t)=2-2X(t-1)$ if $1/2 \leq X(t-1) \leq 1$. For non-linear coupling, $X(t)=2f(t)$ if $0 \leq f(t) < 1/2$ and $X(t)=2-2f(t)$ if $1/2 \leq f(t) \leq 1$, where $f(t)=\epsilon Y(t-1)+(1-\epsilon)X(t-1)$. The strength of coupling, $\epsilon$ is varied from 0 to 0.9 for both simulations. Fig.~\ref{tent_fig} shows the mean values of causality for 50 trials estimated using CCC and TE for linear and non-linear coupling (left and right columns respectively). $N=1000$ (after removal of 1000 initial transient samples), CCC settings used: $L=100$, $w=15$, $\delta=80$, $B=8$. The assumption of a linear model for estimation of GC was proved to be erroneous for most trials and hence GC values are not displayed. As $\epsilon$ is increased for both linear and non-linear coupling, $TE_{Y \rightarrow X}$ increases in the positive direction and then falls to zero when the two series become completely synchronized at $\epsilon=0.5$. The trend of the magnitude of CCC values is similar to TE, however, $CCC_{Y \rightarrow X}$ increment is in negative direction. A non-linear influence of $Y$ on $X$ results in negative causality. 

\textit{Parameter selection for CCC}: Selection of parameters ($L, w, \delta, B$) for the above simulations has been done based on investigations into the nature and computation of Compression-Complexity of time series. The criteria and rationale for the same has been discussed in detail in Section 3 of the Supplemental Material. The same criteria has been applied to the case of real-world data, results of which are discussed below. 

\begin{figure}[t]
 \includegraphics[width=\columnwidth]{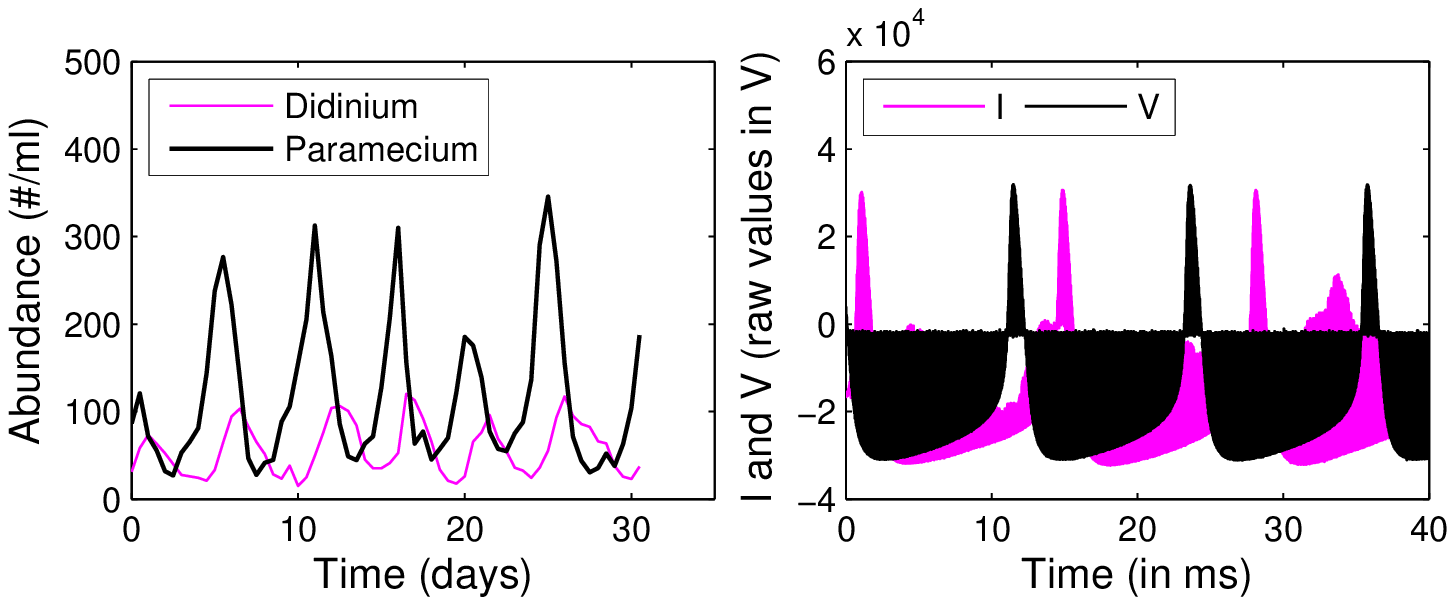}
\hspace*{\fill}
{(a) Predator-Prey system.} \hfill {(b) Squid Axon system.}
\hspace*{\fill}
\vskip 2mm
\centering
\renewcommand{\arraystretch}{1.3}
\newcolumntype{C}[1]{>{\centering\arraybackslash}m{#1}}
\begin{tabular}{|C{1.42cm}|C{1.58cm}|C{1.12cm}|C{1.12cm}|C{1.12cm}|C{1.12cm}|}\hline
\multirow{2}*{\bf System} & \multirow{2}*{\bf Details} & \multicolumn{2}{|C{2.24cm}|}{\bf CCC} & \multicolumn{2}{|C{2.24cm}|}{\bf TE}  \\ \cline{3-6}
& & $Y\rightarrow X$ & $X\rightarrow Y$ & $Y\rightarrow X$ & $X\rightarrow Y$ \\ \hline
Predator- \newline Prey & $Y=D_{n}$ \newline $X=P_{n}$ & 0.116 & -0.021 & 0.969 & 1.077 \\\hline
\multirow{3}*{Squid} & $Y=I_{a3t01}$\newline $X=V_{a3t01}$ & -0.142 & -0.129 & 0.075 & 0.285 \\\cline{2-6}
Axon & $Y=I_{a5t01}$ \newline $X=V_{a5t01}$ & -0.142 & -0.135 & 0.449 & 0.399 \\\cline{2-6}
& $Y=I_{a7t01}$ \newline $X=V_{a7t01}$ & -0.153 & -0.154 & 0.396 & 0.354\\\hline
\end{tabular}
\captionof{figure}{CCC, TE on real-world time series. Top (color online): (a) Time series showing population of \textit{Didinium nasutum} ($D_n$ ) and \textit{Paramecium aurelia} ($P_n$) as reported in \cite{veilleux}, (b) Stimulus current ($I$) and voltage measurements ($V$) as recorded from a Squid Giant Axon (\lq{}a3t01\rq{}) in \cite{paydarfar}. Bottom: Table showing CCC and TE values as estimated for systems (a) and (b).}
\label{real_data_fig}
\end{figure}

\textit{Testing on real-world data}: CCC was applied to estimate causality on measurements from two real-world systems and compared with TE. System (a) comprised of short time series for dynamics of a complex ecosystem, with 71 point recording of predator (Didinium) and prey (Paramecium) populations, reported in \cite{veilleux} and originally acquired for \cite{jost}, with first 9 points from each series removed to eliminate transients (Fig.~\ref{real_data_fig}(a)). $N=62$, CCC settings used: $L=40$, $w=15$, $\delta=4$, $B=8$. CCC is seen to aptly capture the higher (and direct) causal influence from predator to prey population and lower influence in the opposite direction (see Fig.~\ref{real_data_fig}). The latter is expected, owing to the indirect effect of the change in prey population on predator. CCC results are in line with that obtained using CCM~\cite{sugihara}. TE, on the other hand, fails to capture the correct causality direction.

System (b) comprised of raw single-unit neuronal membrane potential recordings ($V$, in 10V) of squid giant axon in response to stimulus current ($I$, in V, 1V=5 $\mu A/ cm^{2}$), recorded in \cite{paydarfar} and made available by \cite{goldberger}. We test for the causation from $I$ to $V$ for three axons (1 trial each) labeled `a3t01', `a5t01' and `a7t01', extracting 5000 points from each recording. $N=5000$, CCC settings used: $L=75$, $w=15$, $\delta=50$, $B=2$. We find that $CCC_{I \rightarrow V}$ is less than or approximately equal to $CCC_{V \rightarrow I}$ and both values are less than zero for the three axons (Fig.~\ref{real_data_fig}), indicating negative causality in both directions. This implies bidirectional non-linear dependence between $I$ and $V$. TE values capture a similar causality magnitude relationship for squid axons `a5t01' and `a7t01', however fails to do so for `a3t01'. 
%
\section{\label{concl}Conclusions and Future Work}
In conclusion, an important contribution of this work is a unifying data compression framework for causality testing which absorbs diverse methods such as GC, TE, CCM, information flow etc. The framework also enabled us to propose a novel \textit{Compression-Complexity Causality} (CCC) measure that outperforms GC and TE for noisy and non-uniformly sampled simulated stochastic data and real-world time series. CCC can be negative, as in the case of linearly and non-linearly coupled chaotic maps, giving rise to the notion of \textit{negative causality}. Such an insightful characterization is absent in existing measures including TE.

Existing methods do not use all the blocks in the framework (Fig.~\ref{block_diagram}). It is hoped that our framework would inspire novel measures other than CCC to be invented in the future. We indicate here a few possibilities. As an example, we have started developing Zip Causality (ZC) testing, based on the popular LZ77 compression algorithm~\cite{ziv} (used by \textit{zip} and \textit{gzip}). ZC measure has all the blocks in the framework except for the quantizer. Our initial testing of ZC measure gave promising preliminary results, but needs further experimentation. Another example of a novel measure would be a \textit{compressed sensing}~\cite{candes} based measure for sparse signals which can efficiently model a wide class of compressible signals. A more futuristic example would be a Deep Learning based causality measure that is also conceivable within our framework. These futuristic measures suggested here are merely illustrative and limited only by our own imagination.

We provide free open access to the CCC \textsc{Matlab} toolbox developed as a part of this work. See Section 4 of Supplemental Material for details. It can be downloaded from the following URL: 
\url{https://sites.google.com/site/nithinnagaraj2/journal/ccc}

\bibliography{main}
\end{document}


\maketitle

In this supplemental material, we provide details of our proposed method Compression-Complexity Causality (CCC) which is not covered in the main paper. We explain the idea of compression-complexity and how it is computed for individual and a pair of time series. We also describe the criteria and rationale for choosing the parameters of CCC and details of our MATLAB implementation that is made available for free download and use.

\section{Compressibility and Compression-Complexity}
There is no single unique definition of ``complexity''. As noted in~\cite{nagaraj3}, Shannon entropy is a very popular and intuitive measure of complexity. A low value of Shannon entropy indicates high redundancy and structure (low complexity) in the data  and a high value indicates low redundancy and high randomness (high complexity). For ergodic sources, owing to Shannon's noiseless source coding theorem~\cite{cover}, (lossless) compressibility of the data is directly related to Shannon entropy. However, robustly estimating compressibility using Shannon entropy for short and noisy time series is a challenge~\cite{nagaraj2}. Recently, the notion of compression-complexity has been introduced~\cite{nagaraj2} to circumvent this problem. Compression-complexity defines the complexity of a time series by using optimal lossless data compression algorithms.  It is well acknowledged that data compression algorithms are not only useful for compression of data for efficient transmission and storage, but also act as models for learning and statistical inference~\cite{Cilibrasi}. Lempel-Ziv  (LZ) Complexity and Effort-To-Compress (ETC) are two measures which fall in this category. 


ETC~\cite{nagaraj1} is defined as the effort to compress the input sequence using the lossless compression algorithm known as Non-sequential Recursive Pair Substitution (NSRPS).  It has been demonstrated that both LZ and ETC outperform Shannon entropy in accurately characterizing the dynamical complexity of both stochastic (Markov) and deterministic chaotic systems in the presence of noise~\cite{nagaraj2,nagaraj3}. Further, ETC has shown to reliably capture complexity of very short time series where even LZ fails~\cite{PEERJ}, and for analyzing short RR tachograms from healthy young and old subjects~{PEERJ}. Thus, we make use of ETC in defining CCC, though it is possible to define CCC using LZ (or other suitable complexity measures).

\begin{table}[!h]
\centering
\caption{Criteria and rationale for choosing the parameters ($w, \delta, B, L$) for CCC. Values of each parameter chosen for Autoregressive (AR), Tent Map (TM), Squid Giant Axon System (SA) and Predator Prey Ecosystem (PP) are enlisted in the rightmost column. Please refer to the main paper for details of these four systems.}
\newcolumntype{C}[1]{>{\centering\let\newline\\\arraybackslash\hspace{0pt}}m{#1}}
\begin{tabular}{|C{0.1\textwidth}|C{0.1\textwidth}|C{0.33\textwidth}|C{0.33\textwidth}|C{0.14\textwidth}|} \hline
\textbf{Parameter} &\textbf{Description}&\textbf{Criteria}&\textbf{Rationale}&\textbf{Values Chosen}\\ 
\hline
$w$ & Window length ${\Delta}X$ & Minimal data length over which CC rate can be reliably estimated. & Earlier studies have revealed that ETC is able to reliably capture complexity of even very short time series~\cite{PEERJ}. & AR: 15 \newline TM: 15 \newline SA: 15 \newline PP: 15 \\ \hline
$\delta$ & Step-size & An overlap of $20-50\%$ between successive time series windows ($X_{past}$ of length $L$) over which CC is estimated. & To capture the continuity of time series dynamics. & AR: 80 \newline TM: 80 \newline SA: 50 \newline PP: 4*  \\ \hline
$B$ & Number of bins & Smallest number of symbols that capture the time series dynamics. & CCC requires symbolic sequences that represent the underlying dynamics. & AR: 2 \newline TM: 8 \newline SA: 2 \newline PP: 8 \\ \hline
$L$ & Window length of immediate past to ${\Delta}X$ ($X_{past}$) & After choosing $w, {\delta}, B$ as above, we plot $ETC(X_{past}+{\Delta}X)$ and $ETC(Y_{past}+{\Delta}X)$ vs. $L$ as well as $ETC(Y_{past}+{\Delta}Y)$ and $ETC(X_{past}+{\Delta}Y)$ vs. $L$ on two separate graphs. \newline \textbf{\textit{First criteria}}: Choose a value of $L$ at which the two curves in both the graphs are well separated. \newline If the above criteria fails (there is an overlap in at least one of the pairs), we plot $ETC(X_{past},Y_{past})$ and $ETC(X_{past}+{\Delta}X,Y_{past}+{\Delta}X)$ vs. $L$ as well as $ETC(Y_{past},X_{past})$ and $ETC(Y_{past}+{\Delta}Y, X_{past}+{\Delta}Y)$ vs. $L$ in separate graphs. \newline \textbf{\textit{Second criteria}}: Choose a value of $L$ such that the two curves in both the graphs are well separated. &  Heterogeneous time series ($Y_{past}+{\Delta}X$) and ($X_{past}+{\Delta}Y$) have very different individual $ETC$ values from that of homogeneous time series ($X_{past}+{\Delta}X$) and ($Y_{past}+{\Delta}Y$) respectively. Due to this clear separation, causation based on CC can be robustly estimated. \newline Heterogeneous time series $(X_{past}+{\Delta}X,Y_{past}+{\Delta}X)$ and $(Y_{past}+{\Delta}Y, X_{past}+{\Delta}Y)$ have very different joint $ETC$ values from that of homogeneous time series $(X_{past}, Y_{past})$ and $(Y_{past},X_{past})$ respectively. Due to this clear separation, causation based on CC can be robustly estimated. & AR: 150 \newline TM: 100 \newline SA: 75 \newline PP: 40 \\ \hline
\end{tabular}
\vspace{0.2in}
*This was an exception with $90\%$ overlap as very short data length was available.
\label{CCC_parameter_criteria}
\end{table}

\section{Individual and Joint Compression Complexities}
In this section, we define how individual and joint compression-complexities are computed using the ETC measure. 

\subsection{ETC measure for a time series: $ETC(X)$}
Since ETC expects a symbolic sequence as its input (of length $> 1$), the given time series should be binned appropriately to generate such a sequence. Once such a symbolic sequence is available, ETC proceeds by parsing the entire sequence (from left to right) to find that pair of symbols in the sequence which has the highest frequency of occurrence. This pair is replaced with a new symbol to create a new symbolic sequence (of shorter length). This procedure is repeated iteratively and terminates only when we end up with a constant sequence (whose entropy is zero since it consists of only one symbol).  Since the length of the output sequence at every iteration decreases, the algorithm will surely halt. The number of iterations needed to convert the input sequence to a constant sequence is defined as the value of ETC complexity. For example, the input sequence `$12121112$' gets transformed as follows: $12121112 \mapsto 33113 \mapsto 4113 \mapsto 513 \mapsto 63 \mapsto 7$. Thus, $ETC(12121112) = 5$. ETC achieves its minimum value ($0$) for a constant sequence and maximum value ($m-1$) for a $m$ length sequence with distinct symbols. Thus, we normalize the ETC complexity value by dividing by $m-1$. Thus normalized $ETC(12121112) = \frac{5}{7}$.  Note that normalized ETC values are always between 0 and 1 with low values indicating low complexity and high values indicating high complexity.

\subsection{Joint ETC measure for a pair of time series: $ETC(X,Y)$}
We perform a straightforward extension of the above mentioned procedure ($ETC(X)$) for computing the joint ETC measure $ETC(X,Y)$ for a pair of input time series $X$ and $Y$ of the same length. At every iteration, the algorithm scans (from left to right) simultaneously $X$ and $Y$ sequences and replaces the most frequent jointly occurring pair with a new symbol for both the pairs.  To illustrate it by an example, consider, $X=121212$ and $Y=abacac$.  The pair $(X,Y)$ gets transformed as follows: $(121212, abacac) \mapsto (1233,abdd) \mapsto (433,edd) \mapsto (53,fd) \mapsto (6,g)$. Thus, $ETC(X,Y)  = 4$ and normalized value is $\frac{4}{5}$. It can be noted that $ETC(X,Y) \leq ETC(X) + ETC(Y)$. 

\section{Parameter selection for CCC: Criteria and Rationale}
In Table~\ref{CCC_parameter_criteria}, we summarize the criteria and rationale for choosing the four parameters ($w, \delta, B, L$) of the proposed measure CCC. These were based on preliminary investigations and explorations into the nature of compression-complexity of various time series. We expect to refine the criteria in the future.

The parameter $w$ which is the length of the moving window ${\Delta}X$ is fixed to 15 for all the datasets used in this work. It is chosen such that it contains sufficient number of data points over which CC rate can be reliably estimated. Earlier studies have revealed that ETC is able to reliably capture complexity of even very short time series (as small as length of 10 samples)~\cite{PEERJ}. $\delta$, the step size by which the ${\Delta}X$ as well as $X_{past}$ window is moved, is chosen based on the criteria of sufficient overlap ($20-50\%$) between successive $X_{past}$ windows of length $L$. $B$, the number of bins used to generate the symbolic sequence of the input time series is chosen such that it is sufficient to capture the underlying dynamics. It was found that for the AR processes, $B \geq 2$ is sufficient whereas the time series from the chaotic tent map requires at least $B=8$.

Once $w, {\delta}, B$ are chosen, we choose $L$, the window length of $X_{past}$. For this, we analyze the curves of ETC measure as it varies with $L$ for different appended (for eg., $X_{past}+\Delta X$, $X_{past}+\Delta Y$) as well as non-appended time series (for eg., $Y_{past}$, $X_{past}$) to estimate their individual (for eg., $ETC(X_{past})$, $ETC(X_{past}+\Delta X)$) as well as joint compression-complexities (for eg., $ETC(X_{past},Y_{past})$, $ETC(X_{past}+\Delta X,Y_{past}+\Delta X)$). A detailed description of selection criteria for $L$ is discussed below.


\begin{figure}[!h]
\includegraphics[width=\textwidth,trim={0 0 0 0},clip]{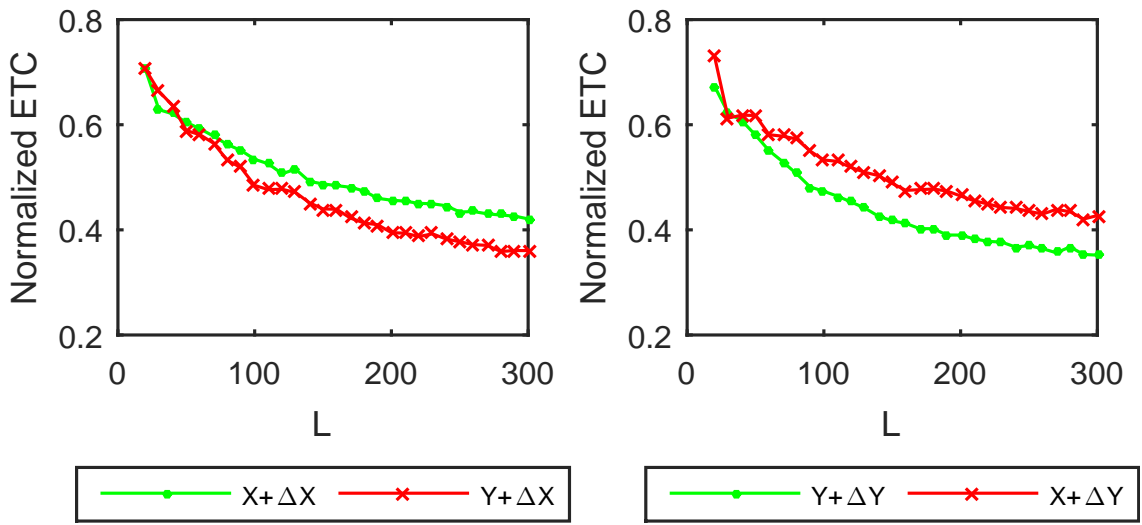}
\caption{(color online). Averaged $ETC(X_{past}+{\Delta}X)$, $ETC(Y_{past}+{\Delta}X)$ curves in the left subfigure and $ETC(Y_{past}+{\Delta}Y)$, $ETC(X_{past}+{\Delta}Y)$ curves in the right subfigure for linearly coupled tent maps ($\epsilon = 0.2$) with $Y$ causing $X$. $w=15, {\delta}=100, B=8$ and $L$ is incremented by a value of 5 data points each time. Using the first criteria for selection of $L$, $L=100$ to $300$.}
\label{ETC_tent_lin_fig}
\end{figure}

\begin{figure}[!h]
\includegraphics[width=\textwidth,trim={0 0 0 0},clip]{ETC_tent_nonlin.eps}
\caption{(color online). Averaged $ETC(X_{past}+{\Delta}X)$, $ETC(Y_{past}+{\Delta}X)$ curves in the left subfigure and $ETC(Y_{past}+{\Delta}Y)$, $ETC(X_{past}+{\Delta}Y)$ curves in the right subfigure for non linearly coupled tent maps ($\epsilon = 0.2$) with $Y$ causing $X$. $w=15, {\delta}=100, B=8$ and $L$ is incremented by a value of 5 data points each time. Using the first criteria for selection of $L$, $L=75$ to $300$.}
\label{ETC_tent_nonlin_fig}
\end{figure}

\begin{figure}[!h]
\includegraphics[width=\textwidth,trim={0 0 0 0},clip]{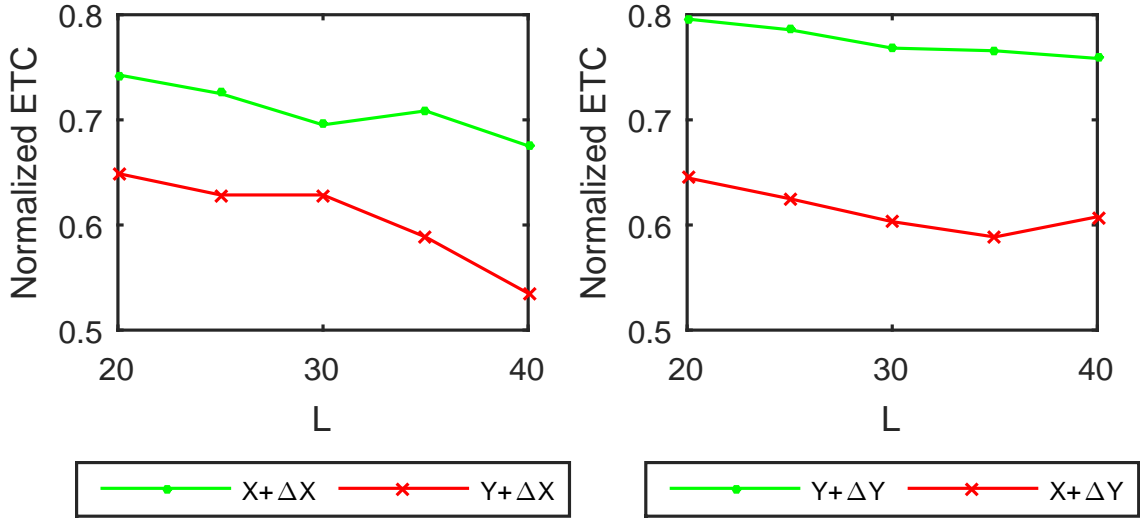}
\caption{(color online). Averaged $ETC(X_{past}+{\Delta}X)$, $ETC(Y_{past}+{\Delta}X)$ curves in the left subfigure and $ETC(Y_{past}+{\Delta}Y)$, $ETC(X_{past}+{\Delta}Y)$ curves in the right subfigure for predator prey ecosystem with $Y$ representing Didinium (predator) population and $X$ representing Paramecium (prey) population. $w=15, {\delta}=1, B=8$ and $L$ is incremented by a value of 5 data points each time. Using the first criteria for selection of $L$, $L=20$ to $40$.}
\label{ETC_preprey_fig}
\end{figure}

\begin{figure}[!h]
\includegraphics[width=\textwidth,trim={0 0 0 0},clip]{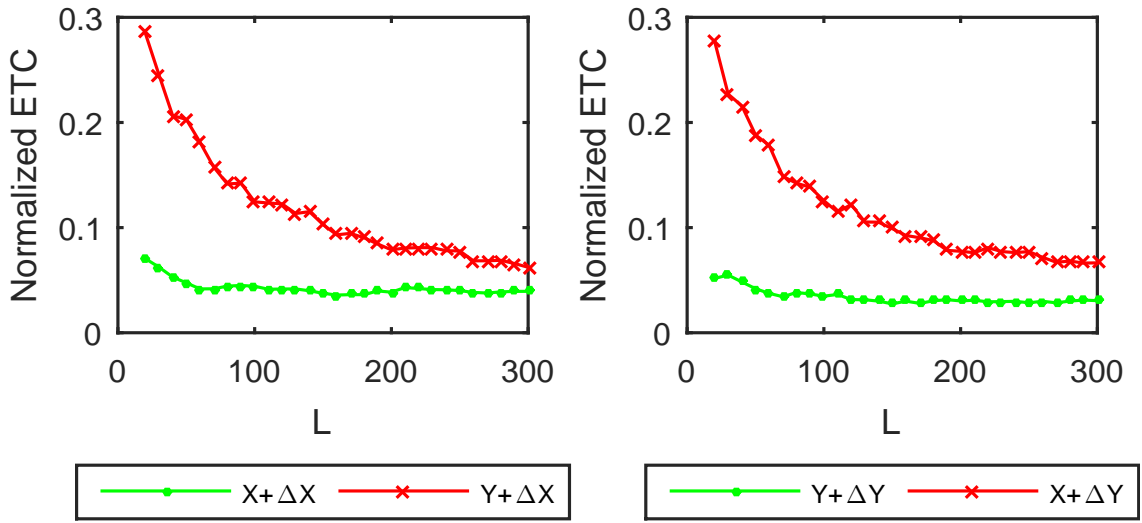}
\caption{(color online). Averaged $ETC(X_{past}+{\Delta}X)$, $ETC(Y_{past}+{\Delta}X)$ curves in the left subfigure and $ETC(Y_{past}+{\Delta}Y)$, $ETC(X_{past}+{\Delta}Y)$ curves in the right subfigure for squid giant axon system (`a5t01') with $Y$ representing the applied stimulus current and $X$ representing observed voltage. $w=15, {\delta}=100, B=2$ and $L$ is incremented by a value of 5 data points each time. Using the first criteria for selection of $L$, $L=75$ to $300$. Lower values of $L$ are not used despite sufficient separation so as to avoid making computation based on the transient stage values.}
\label{ETC_axon5}
\end{figure}

\begin{figure}[!h]
\includegraphics[width=\textwidth,trim={0 0 0 0},clip]{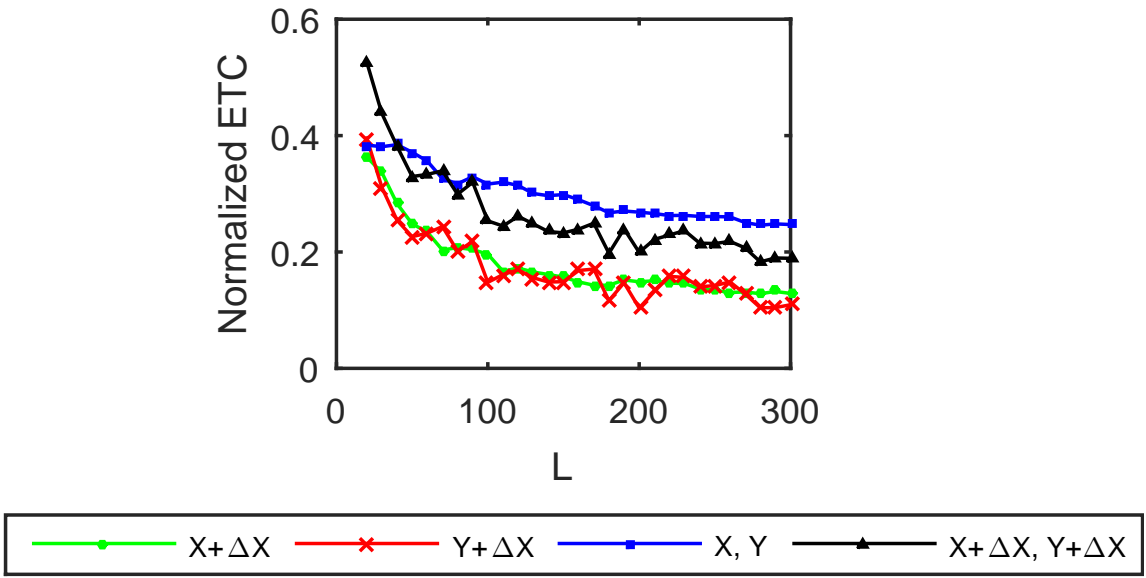}
\caption{(color online). Averaged $ETC(X_{past}+{\Delta}X)$, $ETC(Y_{past}+{\Delta}X)$, $ETC(X_{past}, Y_{past})$, $ETC(X_{past}+{\Delta}X, Y_{past}+{\Delta}X$ curves for coupled AR processes ($\nu = 0.03, \alpha = 0$) with $Y$ causing $X$. $w=15, {\delta}=100, B=2$ and $L$ is incremented by a value of 5 data points each time.}
\label{ETC_AR1_fig}
\end{figure}
\begin{figure}[!h]
\includegraphics[width=\textwidth,trim={0 0 0 0},clip]{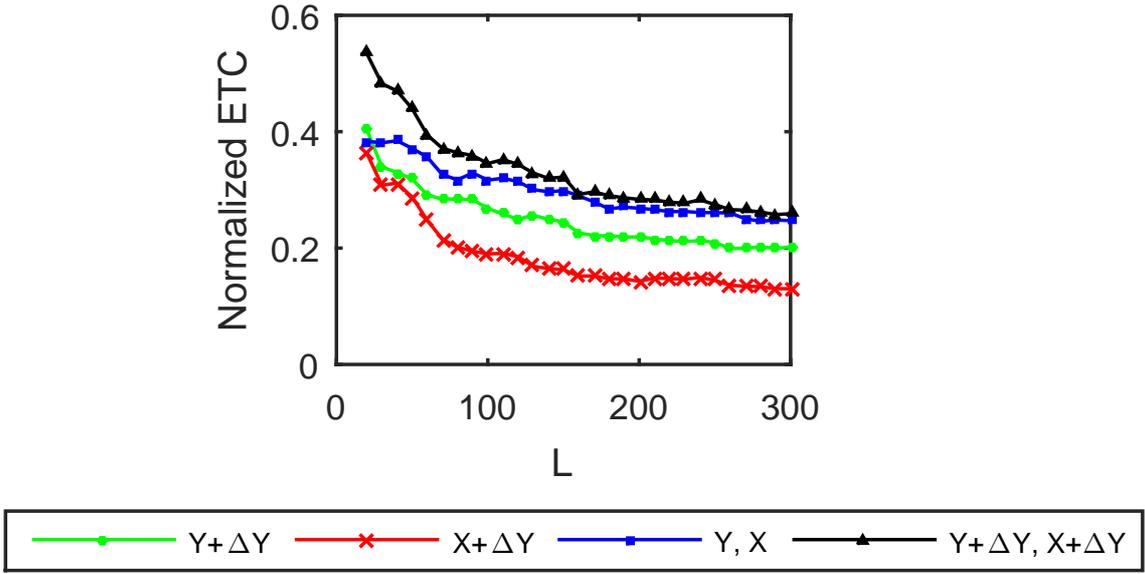}
\caption{(color online). Averaged $ETC(Y_{past}+{\Delta}Y)$, $ETC(X_{past}+{\Delta}Y)$, $ETC(Y_{past}, X_{past})$, $ETC(Y_{past}+{\Delta}Y, X_{past}+{\Delta}Y$ curves for coupled AR processes ($\nu = 0.03, \alpha = 0$) with $Y$ causing $X$. $w=15, {\delta}=100, B=2$ and $L$ is incremented by a value of 5 data points each time. Using the second criteria for selection of $L$, based on this figure and Fig.~\ref{ETC_AR1_fig}, $L=100$ to $160$.}
\label{ETC_AR2_fig}
\end{figure}

\begin{figure}[!h]
\centering
\includegraphics[width=\textwidth,trim={0 0 0 0},clip]{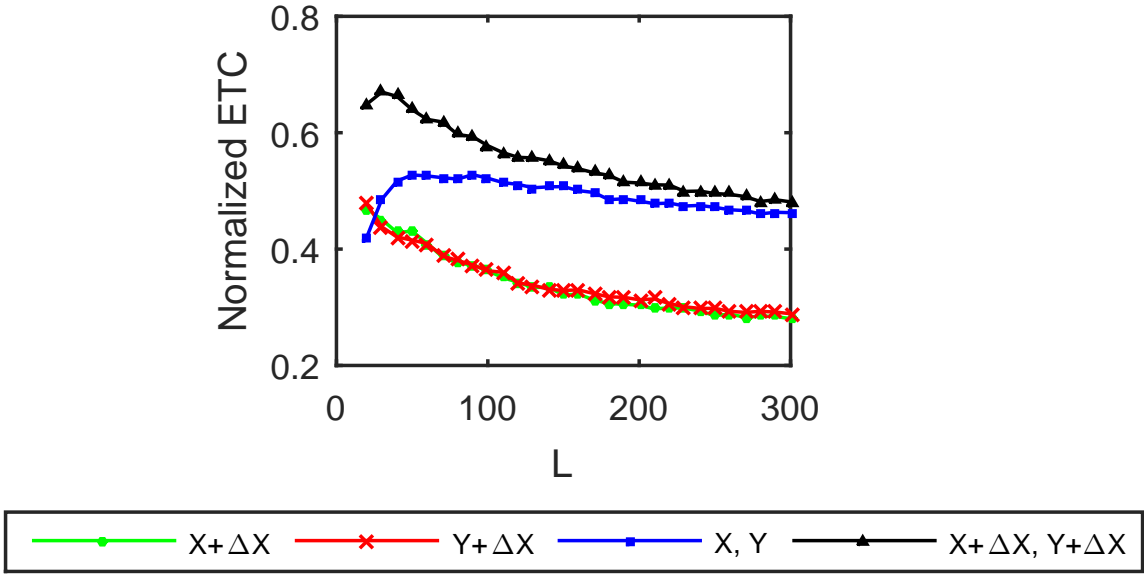}
\caption{(color online). Averaged $ETC(X_{past}+{\Delta}X)$, $ETC(Y_{past}+{\Delta}X)$, $ETC(X_{past}, Y_{past})$, $ETC(X_{past}+{\Delta}X, Y_{past}+{\Delta}X$ curves for independent processes $Y$ and $X$. $w=15, {\delta}=100, B=2$ and $L$ is incremented by a value of 5 data points each time.}
\label{ETC_ind_series}
\end{figure}

\begin{figure}[!h]
\includegraphics[width=\textwidth,trim={0 0 0 0},clip]{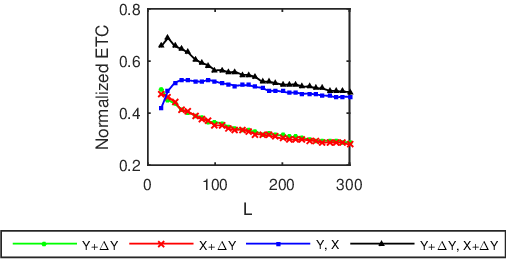}
\caption{(color online). Averaged $ETC(Y_{past}+{\Delta}Y)$, $ETC(X_{past}+{\Delta}Y)$, $ETC(Y_{past}, X_{past})$, $ETC(Y_{past}+{\Delta}Y, X_{past}+{\Delta}Y)$ curves for independent processes $Y$ and $X$. $w=15, {\delta}=100, B=2$ and $L$ is incremented by a value of 5 data points each time. Using the second criteria for selection of $L$, based on this figure and Fig.~\ref{ETC_ind_series}, $L=100$ to $300$, avoiding the range of $L$ giving transient values of CCC.}
\label{ETC_ind_series_opp}
\end{figure}

\subsection{Selection Criteria for $L$}


As discussed in Table~\ref{CCC_parameter_criteria}, for given time series $X$ and $Y$, we first plot $ETC(X_{past}+{\Delta}X)$ and $ETC(Y_{past}+{\Delta}X)$ vs. $L$ as well as $ETC(Y_{past}+{\Delta}Y)$ and $ETC(X_{past}+{\Delta}Y)$ vs. $L$ on two separate graphs. We choose a value of $L$ at which the two curves in both the graphs are well separated. In this work, we start with an $L=20 (> w)$ and go up to $L=300$ (in case of the predator prey ecosystem data, only 62 data points were available and thus we go up to $L=40$). In Figs.~\ref{ETC_tent_lin_fig}, \ref{ETC_tent_nonlin_fig}, \ref{ETC_preprey_fig} and \ref{ETC_axon5} which show these curves plotted for linearly and non-linearly coupled tent maps, predator prey and squid giant axon systems respectively, there exists some range of values of $L$ for which the two curves are well separated. A value of $L$ can thus be chosen from within this range.

The separation between the curves exhibits that heterogeneous time series ($Y_{past}+{\Delta}X$) and ($X_{past}+{\Delta}Y$) have very different ETC values from that of homogeneous time series ($X_{past}+{\Delta}X$) and ($Y_{past}+{\Delta}Y$) respectively. A clear separation between these two curves cannot be merely accounted by statistical variations and most likely can be traced to some kind of causal relationship.

If the above criteria fails (there is an overlap in at least one of the pairs), we plot $ETC(X_{past},Y_{past})$ and $ETC(X_{past}+{\Delta}X,Y_{past}+{\Delta}X)$ vs. $L$ as well as $ETC(Y_{past},X_{past})$ and $ETC(Y_{past}+{\Delta}Y, X_{past}+{\Delta}Y)$ vs. $L$ in separate graphs. We choose a value of $L$ such that the two curves in both the graphs are well separated. In case of AR processes where the first criteria is not met due to the overlap between $ETC(X_{past}+{\Delta}X)$ and $ETC(Y_{past}+{\Delta}X)$, the second pair of curves is plotted as shown in Figs.~\ref{ETC_AR1_fig} and \ref{ETC_AR2_fig}. The rationale behind this criteria is similar to that encountered in the case of first criteria. Here, the idea is to ensure that joint complexities  of heterogeneous time series $(X_{past}+{\Delta}X,Y_{past}+{\Delta}X)$ and $(Y_{past}+{\Delta}Y, X_{past}+{\Delta}Y)$ have very different ETC values from homogeneous time series $(X_{past}, Y_{past})$ and $(Y_{past},X_{past})$ respectively. Given the separation, we can safely compute their differences for CC rate estimation with conviction that the difference is not arising due to statistical differences.

We have, as yet, not encountered a case where both the above criteria fail. We are not sure whether this is plausible. Even in the case of two independent and uniformly distributed real time series, though the first criteria fails, the second one is valid (Figs.~\ref{ETC_ind_series}, \ref{ETC_ind_series_opp}). Any value of $L$ used from the estimated range to compute CCC will result in a value which is statistically close to zero in both directions.

\section{Description of CCC Toolbox}

The accompanying CCC toolbox, implemented in MATLAB contains the following files:

\begin{enumerate}
\item \textbf{Main.m} calls functions to simulate coupled AR processes or tent maps and to estimate the value of Compression-Complexity Causality between them.
\item \textbf{coupled\_AR.m} simulates coupled AR processes with a desired level of noise or percentage of non-uniform sampling.
\item \textbf{puncture.m} introduces non-uniform sampling/non-synchronous measurements in the data.
\item \textbf{coupled\_tent.m} simulates non-linearly coupled tent maps.
\item \textbf{UpdateTent.m} updates the values of the tent map at every iteration.
\item \textbf{CCC.m} estimates Compression-Complexity Causality between two input time series.
\item \textbf{ETC\_1D.m} and \textbf{ETC\_2D.m} subroutines estimate individual and joint ETC values respectively.
\item \textbf{Partition.m} bins the given time series before estimating ETC values.
\end{enumerate}

\bibliographystyle{ieeetr}
\bibliography{supp_references.bib}